\documentclass[prd,preprint,tightenlines,floatfix,showpacs,preprintnumbers,nofootinbib,eqsecnum]{revtex4}

 \usepackage[dvips,final]{graphicx}
  \usepackage{amssymb}
   \usepackage{amsmath}
    \usepackage{amsfonts}
     \usepackage{epsfig}
      \usepackage{bm}% bold math
\usepackage[english]{babel}

\numberwithin{equation}{section}

\begin{document}

\title{The gauge model of quark--meson interactions and its application to the meson radiative decays}% Force line breaks with \\

\author{V.~Beylin}
\email{vbey@rambler.ru}
\author{V.~Kuksa}%
 \email{kuksa@list.ru}
\author{G.~Vereshkov}
\email{gveresh@gmail.com} \affiliation{Research Institute of
Physics, Southern Federal University, Rostov-on-Don 344090,
Russia}
\date{\today}

\begin{abstract}

To analyze an electromagnetic and strong hadron processes at low energies, we consider the renormalizable model with the
$U_0(1)\times U(1)\times SU(2)$ gauge symmetry. This approach is
based on the linear sigma-model extended by the gauge and
quark-meson interactions. Physical content and parameters of the
model are discussed. Theoretical predictions for some radiative
decays of vector mesons are in a good agreement with the
experimental data.

\end{abstract}


\pagenumbering{arabic}\setcounter{page}{1}

\maketitle

\section{Introduction}
The low-energy processes with a hadron participation are
permanently in the center of theoretical and experimental
activity. New measurements and more precise experimental data
force us to look for and verify new theoretical approaches to describe and understand strong coupling phenomena.

It is well known, an exact theory of (low-energy) hadron
interactions should be nonperturbative (NP). That is why a lot of
hadron characteristics (effective couplings, formfactors etc.) are
calculated at the level of the fundamental QCD fields, which are submerged
into the NP vacuum. In particular,
 QCD Sum Rules method operates with some phenomenological NP vacuum parameters
 - non-zero quark and gluon condensates - to simulate manifestations of the vacuum and its affection on hadron characteristics.
 Effective theories of meson-meson interactions provide an additional
 possibility to extract information on the low-energy hadron dynamics and structure, accounting
 for the NP vacuum via phenomenological parameters.

There are two well known and widely used methods of effective
Lagrangian's approach in hadron physics: derivation of the effective Lagrangian
from the QCD immediately \cite{1}-\cite{11e},
and the use of various dynamical symmetries to build the basic
Lagrangian in a phenomenological way \cite{12}-\cite{18}. In
particular, the gauge scheme of vector meson interactions with
hadrons was suggested in \cite{15}-\cite{17}. The linear
sigma model $(L\sigma M)$ is the most popular and examined
approach to the hadron phenomenology, which is used for the
effective study of nucleon-nucleon \cite{19}-\cite{21} and
quark-meson interaction \cite{12},\cite{13},\cite{22}-\cite{24}
with regard to the vector meson dominance \cite{25}-\cite{27}.

When the fundamental QCD is a starting point, chiral $SU_L(2)\times SU_R(2)$ perturbation theory
\cite{11a} follows from a low-energy expansion
of the QCD Green functions in powers of the external momenta and of the (low) quark current masses.
This theory leads to an effective low-energy Lagrangian, which includes some phenomenological
coupling constants and allows to analyze $\pi \pi$ - scattering and pion formfactor as well.
Generalization of the theory with $U_L(3)\times U_R(3)$ symmetry has been done, and it was used,
especially, for the calculation of the masses and decay constants for the pseudoscalar octet \cite{11a} -\cite{11e}.

The chiral model can be also deduced from
the interacting QCD fields due to some
bosonization procedure, transferring the gauge ideas from the quark level QFT to the
meson one \cite{4} in the framework of the Nambu-Jona-Lasinio-type Lagrangian with the six-quark interaction, suggested by t'Hooft.
In such a way, initial quarks with low current masses transform to the
constituent ones with effective masses $\sim 300 \,MeV$, including
some contribution of the internal gluon substructure of a hadron.

The sigma model at the quark-meson level can be dynamically generated via (diverged)
quark loops with a cutoff parameter \cite{27a}, \cite{27b}.
Therewith, the loops simulate an uprising of the
nonperturbative model "`condensate". So, the proportionality of pion
mass to the "quark condensate" is restored in the framework of
such effective models. Further, to incorporate
electromagnetic and strong vectorlike interactions into the
effective theory an appropriate vector fields are introduced as
the gauge fields. Treatment of the vector mesons as the gauge fields,
which realize a dynamical symmetry, allows to diminish some
theoretical uncertainty of phenomenological description of the
hadron interactions.

Effective gauge models can be formulated as the phenomenological ones, based on the specific dynamical symmetry of interactions. For example, in ~\cite{15}-\cite{18} the gauge models were successfully
used for the consideration of some low-energy aspects of
baryon-meson interaction. It means that the fundamental quantum
field principles (the gauge symmetry) can be applied to the description of interactions
at the different hierarchical levels. In the case under
consideration, it provides a transition from the quark-level sigma
model ($Q \sigma M$) to the nucleon-level sigma model ($N\sigma
M$) (as it was shown in \cite{3a}, \cite{3c}, \cite{4} and \cite{11a}, \cite{11b}, the nonlinear sigma model structure can be derived from the QCD).

To construct an effective meson interaction model, we use the
gauge scheme based on $U_0(1)\times U(1)\times SU(2)$ group. This
group is the simplest one to consider the light unflavored vector
mesons, $\rho$ and $\omega$, together with the photon, as the
gauge fields. Moreover, it has the necessary symmetry to study processes with the vector fields
that are mediated by the electromagnetic and strong meson-meson and quark-meson
interactions. Here, we
consider only the interactions that are insensitive to the chiral
structure (the gauge generalization of corresponding chiral group is in progress).

 For the case in question, to analyze the
strong effects it is sufficient to localize $SU(2)$ group - the diagonal sum of
chiral $SU_{L,R}(2)$ subgroups of the global
$SU_L(2)\times SU_R(2)$. The extra $U(1)$ groups are added
for the gauge realization of the vector meson dominance (VMD)
together with the electromagnetic interaction (see the next
section). It is known, the gauge scheme of VMD is provided by the
electromagnetic quanta mixing with $\omega$- and $\rho$- fields.
In particular, this scheme describes some important features of
radiative decays of vector mesons.

After the spontaneous breaking of the $U_0(1)\times U(1)\times
SU(2)$ local symmetry, residual scalar massive degrees of freedom
(Higgs fields) should be interpreted as some observable scalar
mesons. More exactly, they can be associated with the scalar
isotriplet $a_0(980)$ and isosinglet $f_0(980)$ (see the next section).

This model reproduces the relation $m^2_{\pi}\sim m_q <\bar q q>$
if the quark condensate is taken into account in the equation for
the vacuum shift. Here, we deal with the constituent quarks not
with the fundamental current ones. These effective fields in the
model are taken as the fundamental representation of the gauge
group. Then, (effective) quarks degrees of freedom are used to account for the effects of
an internal hadron structure in meson-meson interactions. In fact, quark-meson models (see also
\cite{22}, \cite{27b}, \cite{28}, \cite{4}, \cite{11a}, \cite{11b}) are necessary hybrid approaches to reproduce both
nonperturbative and structure effects of the low-energy meson
interactions. Manifestations of the fundamental hadron components are reproduced by the constituent quarks and mesons, $\sigma$ and $\pi$. Then, the last ones have a dual role in the model - they appear both as an external fields (particles of matter) and as an effective virtual components of internal hadron vacuum: "gluon" degrees of freedom ($\sigma$-meson) and an excitation of the quark sea ($\pi$-meson, which is linked with the quark current).

So, we deal with the $\sigma$ model inspired
consideration, which incorporates gauge interactions of vector mesons
with the constituent quarks. This quantum field model is renormalizable "by the construction" and it accounts for hadronic
internal degrees of freedom as a quantum
fields. In essence, two-level (quark-meson) structure of the
model realizes an analog of the "bootstrap" idea.

In this work processes of the vector meson decays are only considered.
The gauge sector of the model contains a number of free parameters, which can be fixed from the well measured two-particle
decays and masses of vector mesons. As it will be shown later, the
gauge model predictions for the radiative vector meson decays do not contain any free parameters.

As a rule, the radiative decays $\rho^0 \to \pi^+ \pi^-\gamma$ and
$\omega \to \pi^+ \pi^-\gamma$ are described in a phenomenological
way. Within the framework of this gauge model, these reactions
arise at the tree level. As it will be shown, there is a good agreement between the
theory and the experimental data on the vector meson radiative
decays even without loop corrections $\sim 1/N_c$ (see also \cite{28a} and the Section 3). So, from the direct calculations it follows that
these processes can be successfully studied in the gauge
field approach.

Main contribution of an internal hadron structure into physical processes is described by the quark-meson sector of
the model. This aspect of the model has been tested in $\pi^0 \to \gamma \gamma$, $\rho
\to \pi^0 \gamma$, $\omega \to \pi^0 \gamma$ decays, which arise due to the quark-meson interaction at the loop
level only. The results of calculations are in a good agreement
with the experimental data. Analysis of the decay $\omega (\rho) \to
\pi^0 \pi^+ \pi^-$ \cite{28},\cite{29a}-\cite{29c} is much more cumbersome.
Calculations done show that the three-pion decays of $\omega-$ and
$\rho-$ mesons are well described in this model. The results
obtained are in preparation now.

The structure of the paper is as follows. In Section 2 the gauge
field model is described, and some details of calculational procedure are also presented. Numerical results for tree decays
$\rho^0\rightarrow\pi^+\pi^-\gamma$ and
$\omega\rightarrow\pi^+\pi^-\gamma$ are discussed in Section 3.
We consider loop decays $\rho^0\rightarrow\pi^0 \gamma$, $\omega\rightarrow\pi^0 \gamma$
and $\pi^0\to \gamma \gamma$ in Section 4.

\section{The gauge field model of meson-meson\\ and quark-meson interactions}

The gauge model of the low-energy meson interactions is based on
the conceptual assumption concerning the transfer of gauge
principles from the fundamental quark level to the effective
hadron one. An explicit transfer via the bosonization procedure
was described, for example, in \cite{4} for the case of quark-meson hierarchy
levels. Remind also, meson physics follow from the QCD through low-energy expansion of quark Green functions in the chiral form \cite{11a}, \cite{11b}.
So, an effective quark-meson theories can be derived from the "first principles".
 Applicability of the gauge approach to the
baryon-meson interaction has been demonstrated in \cite{15}-\cite{18}. In this work, we show the validity of the
gauge scheme with the spontaneous symmetry breaking for the studying of meson-meson and quark-meson
effective interactions together with the vector meson dominance.

Strong and electromagnetic interactions are united in the gauge
scheme, and the simplest variant of corresponding dynamical
symmetry is based on $U_{0}(1)\times U(1)\times SU(2)$ group. In a
general case, the total chiral group of symmetry should be
localized to consider both vector and axial-vector mesons.
However, for the application of the gauge approach to the $\rho-$
and $\omega-$ strong and electromagnetic decays only, it is
sufficiently to localize the $SU(2)$ group together with the
additional $U(1)$ groups.

The $\sigma$- model symmetry can be realized in explicitly chiral
form as the representation of $SU_L(2)\times SU_R(2)$ global group. Then,
the quark doublet transformation is:
$$
q^{'}=q+\frac{i}{2}(\alpha+\beta)_a\tau^a
q_R+\frac{i}{2}(\alpha-\beta)_a\tau^a
   q_L.
$$

This transformation can be also rewritten in an equivalent
form with the same number of independent parameters, $\alpha_a$
and $\beta_a$:
\begin{align}\label{2.0}
q^{'}&=q+\frac{i}{2}(\alpha_a\tau^a+\beta_a\tau^a\gamma_5)q.
\end{align}

Having regard to the standard structure of the quark interaction with $\sigma$ and $\pi$ (see the second term in the (\ref{E:8})),
parameters $\alpha_a$
and $\beta_a$ define transformation properties of $(\sigma,\pi)$
multiplet with respect to the global group:
\begin{align}
\pi^{'}_a=&\pi_a+\epsilon_{abc}\alpha_b \pi_c+\beta_a\sigma,\notag\\
\sigma^{'}=&\sigma+\beta_a\pi_a\notag.
\end{align}

As it is seen from the expressions above, the set of $\alpha_a$
parameters corresponds to the diagonal sum of $SU_{L,R}(2)$
subgroups, i.e. the $SU(2)$ group. In the model, just this group is
localized to analyze strong interactions of $\rho-$ meson triplet.
To introduce $\omega$- singlet and electromagnetic field, we add
two extra $U(1)$ groups, which should be localized too. Certainly,
if the total chiral group were localized, axial vector gauge
fields will occur as a chiral partners of vector mesons. However,
to describe radiative processes with vector mesons only we
restrict ourselves by the localization of above mentioned groups.

Here, the triplet
$\pi_a$, the singlet $\sigma$-meson and $u, d$-quarks are the
initial fields of matter, where the triplet of pions is the adjoined representation
of the $SU(2)$ gauge group, and the quark doublet, $u, \,d$, is
the fundamental representation of this one. Remind, the $\sigma$ and $\pi$ have a dual nature in the model both due to their contribution to the processes inside a hadron NP medium and their manifestations as the real mesons.

Note, also, that the $\sigma$ field does not related directly with the non-singlet (flavor) quark current. So, we can assume the
$\sigma-$ meson is a "gluonium" state, and it can be interpreted
as $f_0(600)-$ meson \cite{22}, \cite{29pa}-\cite{29cd}. At the
same time, physical massive scalar fields, emerging from the
initial Higgs multiplets, can be associated with the scalar mesons:
isotriplet $a_0(980)$ and isosinglet $f_0(980)$ (see below). Naturally, this
interpretation should be confirmed by the study of the scalar sector.
The detailed analysis of the scalar mesons properties (masses, couplings, decay widths) in the gauge model will be done in a forthcoming paper.

From these considerations with an account of the self-action
nonlinear terms,
 the initial model Lagrangian can be written as follows:
\begin{align}\label{E:8}
L &=i \bar{q} \hat{D} q - \varkappa \bar{q} (\sigma +i\pi^a\tau_a \gamma_5)q +
\frac{1}{2}(D_{\mu}\pi^a)^+(D_{\mu}\pi^a)+\frac{1}{2}\partial_{\mu}\sigma
\partial^{\mu}\sigma + \frac{1}{2}\mu^2(\sigma^2+\pi^a\pi^a)\notag \\
&-\frac{1}{4}\lambda(\sigma^2+\pi^a\pi^a)^2+(D_{\mu}H_A)^+(D_{\mu}H_A)+
\mu_A^2(H_A^+H_A)-\lambda_1(H_A^+H_A)^2\notag\\
&-\lambda_2(H_A^+H_B)(H_B^+H_A)
-h(H_A^+H_A)(\sigma^2+\pi^a\pi^a)-\frac{1}{4}B_{\mu \nu}B^{\mu
\nu}\notag\\
&-\frac{1}{4}V_{\mu \nu}V^{\mu \nu}-\frac{1}{4}V^a_{\mu
\nu}V_a^{\mu \nu}.
\end{align}
Here $q = (u,d)$ - is the first generation quark doublet;
$H_{1,2}$ - two doublets of complex scalar fields with the
hypercharges $Y_{1,2}=\pm 1/2$, $a=1,2,3$ and $A=1,2$. The gauge
derivatives and field strengthes are:
\begin{align}
&\hat{D}q = \gamma^{\mu}(\partial_{\mu}-
\frac{i}{6}g_{0}B_{\mu}-\frac{i}{2}g_1 V_{\mu}-\frac{i}{2}g_2 V^a_{\mu}\tau_a)q;\notag\\ &D_{\mu}\pi_a=\partial_{\mu}\pi_a-ig_2 V^b_{\mu}\epsilon_{bac}\pi_c;\notag\\
&D_{\mu}H_{1,2} = (\partial_{\mu}\pm
\frac{i}{2}g_{0}B_{\mu}-\frac{i}{2}g_1 V_{\mu}-\frac{i}{2}g_2\tau_a
V^a_{\mu})H_{1,2};\notag\\
&B_{\mu\nu}=\partial_{\mu}B_{\nu}-\partial_{\nu}B_{\mu},\,\,\,V_{\mu\nu}=
\partial_{\mu}V_{\nu}-\partial_{\nu}V_{\mu};\notag\\
&V^a_{\mu\nu}=\partial_{\mu}V^a_{\nu}-\partial_{\nu}V^a_{\mu}+g_2\epsilon ^{abc}V^b_{\mu}V^c_{\nu}.
\label{E:9}
\end{align}
In principle, the gauge invariant term $B_{\mu \nu}V^{\mu \nu}$
should be added to the Lagrangian. However, this term can be
diagonalized from the very beginning by the orthogonal
transformation of the initial $B_{\mu}$ and $V_{\mu}$ fields. So,
we get the same Lagrangian with the redefined $B_{\mu}$ and
$V_{\mu}$ fields.

There is one further consequence of the $B_{\mu \nu}V^{\mu \nu}$
term diagonalization. Namely, to describe the realistic processes
of decays and the lepton scattering we can introduce interactions
of leptons with $U_{0}(1)$ and $U(1)$ fields in the gauge  form:
\begin{equation}
L_l=i\bar l \hat D l=i \bar l \gamma^{\mu}(\partial _{\mu} -
ig_{0}B_{\mu} -i\varepsilon g_1V_{\mu})l,
\label{E:10}
\end{equation}
where interaction of $V_{\mu}$ field with leptons is driven by the
phenomenological coefficient $\varepsilon$. The last term in the above
 formula is just the coupling, which is resulted from the diagonalization of
 $B_{\mu \nu}V^{\mu \nu}$ term
in the initial Lagrangian. Thus, the above interaction of leptons with
the $V_{\mu}$ vector field can be used to describe the
decays $\rho^0\to e^{+}e^{-},\,\,\,\omega\to e^{+}e^{-}$. Analogous point-like interaction of lepton doublets
with scalar doublets is forbidden in the model by the symmetry and
renormalizability demands.

Physical states are formed by the primary fields mixing when
quadratic mass forms of scalar and vector fields are diagonalized.
At the tree level, the mass forms arise as a result of vacuum
shifts:
\begin{equation}
<\sigma>=v,\,\,\, <H_1>=\frac{1}{\sqrt{2}}(v_1,0),\,\,\,
<H_2>=\frac{1}{\sqrt{2}}(0,v_2). \label{10a}
\end{equation}
This shift defines the gauge fields masses at the tree level entirely.
%In a general case, the mass matrix has the form:
%$$M^2(s)=M^2_0+\Pi(s),$$
%where $M^2_0$ is formed by the shifts (\ref{10a}) and it
%corresponds to unphysical fields yet. An account of the
%self-energy contribution $\Pi(s)$ leads to the physical mass
%spectrum. Namely, diagonalizing the total mass forms $M^2(s)$ we
%get the model spectrum of the vector and scalar multiplets.
%Moreover, the self-energy insertions are significant for the
%description of the $\omega-\rho^0$ mixing (see a comment below).

It is known, the physical photon has a hadron (quark) component, which manifests itself in interactions with meson fields.
In the model, an emergence of this primary hadron component in the photon results from the mixing in the gauge sector. The
structure of the vector boson physical states can be illustrated
in a tree approximation (when the mass matrix $M^2$ does not contain any loop corrections) by the following
expressions:
\begin{align}\label{E:11}
&A_{\mu}=\cos\theta \cdot B_{\mu}+\sin\theta \cdot V^3_{\mu}, \notag\\
&\omega_{\mu} = \cos\phi \cdot V_{\mu}+\sin\phi \cdot
(\sin\theta\cdot B_{\mu}-\cos\theta \cdot V^3_{\mu}),\notag\\
&\rho^0_{\mu} = \sin\phi \cdot V_{\mu}+\cos\phi \cdot
(-\sin\theta\cdot B_{\mu}+\cos\theta \cdot V^3_{\mu}),
\end{align}
where the mixing angle $\theta$ is determined by the
diagonalization of the vector fields quadratic form. Due to the mixing
(\ref{E:11}) the processes with initial "unphysical" photon,
$e^+e^-\to \gamma '\to X$, receive contributions from intermediate hadron states:
\begin{equation}\label{E:12}
e^+e^-\to \gamma,\omega, \rho^0\to X.
\end{equation}
 Some parameters of the mixing can be fixed from the experimental data on the vector meson masses and decay widths.
 It will allows to describe decay properties of vector mesons, which are
considered in this paper.
As it can be seen from the following, description of vector mesons (their radiative decays, in particular) in this gauge scheme
is in a good agreement with the experimental data \cite{29d}, confirming the gauge status of
$\rho(770),\,\,\omega(782)$ and photon.

Now, we give the main part of the physical Lagrangian which will be used for calculations:
\begin{align}\label{E:12a}
 L_{Phys}&=\bar{u}\gamma^{\mu}u(\frac{2}{3}eA_{\mu}+g_{u\omega}\omega_{\mu}+g_{u\rho}\rho^0_{\mu})+
          \bar{d}\gamma^{\mu}d(-\frac{1}{3}eA_{\mu}+g_{d\omega}\omega_{\mu}+g_{d\rho}\rho^0_{\mu})\notag\\
          &+ig_2(\pi^{-}\pi^{+}_{,\mu}-\pi^{+}\pi^{-}_{,\mu})(\sin \theta \,\, A^{\mu}-\cos \theta \sin \phi \,\,
         \omega^{\mu}+\cos \theta \cos\phi \,\,\rho^{0\mu})\notag\\
         &-\sqrt{2}i\varkappa\pi^{+}\bar{u}\gamma_5 d-\sqrt{2}i\varkappa\pi^{-}\bar{d}\gamma_5 u-i\varkappa\pi^0
         (\bar{u}\gamma_5 u-\bar{d}\gamma_5 d)\notag\\
         &+2g_2e\cos \theta \cos\phi \,\,\rho^0_{\mu}A^{\mu}\pi^{+}\pi^{-}-
          2g_2e\cos \theta \sin\phi
          \,\,\omega_{\mu}A^{\mu}\pi^{+}\pi^{-}\notag \\
          &+\frac{1}{\sqrt{2}}g_2 \rho^{+}_{\mu}\bar{u}\gamma^{\mu}d+\frac{1}{\sqrt{2}}g_2 \rho^{-}_{\mu}\bar{d}
         \gamma^{\mu}u
          +ig_2\rho^{+\mu}(\pi^0\pi^{-}_{,\mu}-\pi^{-}\pi^0_{,\mu})\notag \\
          &+ig_2\rho^{-\mu}(\pi^{+}\pi^0_{,\mu}-\pi^0\pi^{+}_{,\mu}).
\end{align}
In Eqs.(\ref{E:12a}):
\begin{align}\label{E:12b}
 &g_{u\omega}=\frac{1}{2}g_1 \cos\phi +\frac{1}{2}\sin\phi \,(\frac{1}{3}g_0\sin \theta-g_2\cos \theta)  \,,\notag\\
 &g_{u\rho}=\frac{1}{2}g_1 \sin\phi -\frac{1}{2}\cos\phi \,(\frac{1}{3}g_0\sin \theta-g_2\cos \theta)\,, \notag\\
 &g_{d\omega}=\frac{1}{2}g_1 \cos\phi +\frac{1}{2} \sin\phi \,(\frac{1}{3}g_0\sin \theta+g_2\cos \theta) \,,\notag\\
 &g_{d\rho}=\frac{1}{2}g_1 \sin\phi -\frac{1}{2} \cos\phi \,(\frac{1}{3}g_0\sin \theta+g_2\cos
 \theta)\,.
\end{align}
Here $\sin{\phi}, \,\, \cos{\phi}$ - are tree
mixing parameters (see (\ref{E:11})). Note also, from the point of view of this mixing $\omega$- meson is not a pure isoscalar state in the model.
However, the isotriplet admixture to the $\omega$ structure is very small:
it is proportional to $\sin\phi$
which is close to zero, as we will see later. So, the corresponding contribution can be omitted in calculations involving the $\omega$- meson in a good approximation.

%%They become a complex
%(renormalized) parameters of the $V-B$ mixing after an account of
%the self-energy insertions to the mass matrix. In this case, the
%parameters (\ref{E:12a}) should be treated as an effective complex
%couplings. It is an essential detail to describe the $\omega -
%%\rho$ interference.

As a direct consequences of the
model, we have the following relations:
 \begin{equation}\label{E:12c}
  \sin \theta=\frac{g_{0}}{\sqrt{g_0^2+g_2^2}},\,\,\,e=g_0\cos \theta,\,\,\, v^2_1+v^2_2=4 \frac{m^2_{\rho^{\pm}}}
  {g^2_2},\,\,\,
  \sin\phi=\frac{g_1}{g_2}\Bigl(\frac{m^2_{\rho^{\pm}}-m^2_{\omega}(g^2_2/g^2_1)}
  {m^2_{\omega}-m^2_{\rho^0}}\Bigr)^{1/2}.
 \end{equation}

Introduction of the vector fields into the theory in a gauge way
provides universality of couplings with the vector fields. An
analogous universality of the vector and pseudoscalar meson
interactions was analyzed in \cite{24} in a phenomenological way.
This universality strictly bounds the number of free parameters in
the gauge sector of the model.
Namely, from (\ref{E:12a}) it follows that the vector boson interactions with mesons is described by three gauge parameters: $g_{0}$, $g_{1}$ and $g_{2}$.

The way to analyze low-energy quark-meson Lagrangians is well known: effective couplings of vector boson interaction with mesons and (constituent) quarks are defined by the (diverged) quark loops. In such a way the divergencies are regularized with a cutoff parameter $\Lambda$, which, in fact, fixes some NP scale of the theory. An analogous scheme can be applied to the physical Lagrangian (\ref{E:12a}). Namely, due to renormalizability of the model an effective couplings can be reproduced by the sum of (bare) tree and one-loop quark and meson (logarithmically diverged) diagrams, which also introduce some NP scale ($\Lambda \sim 1 \,\mbox{GeV}$). Certainly, to define the effective couplings in this model we should add to the tree vertices all extra loops with scalars (i.e. not only $\pi \sigma \pi$ - triangles as in \cite{22}, \cite{24}, \cite{27a}, \cite{27b}) and to take into account an existence of the mixing angles, $\phi$ and $\theta$, in the couplings.
Then, the sums of (bare) tree vertices in (\ref{E:12a}) and (logarithmically diverged) loop corrections are interpreted as the effective phenomenological couplings on the mass shell of external particles. These couplings incorporate both short-distance and long-distance contributions from the very beginning and their values can be fixed from the data on meson two-particle decays with a sufficient accuracy. We call them later as the gauge "tree" (on-shell) vertices having the bare vertices tensor structure.

Then, we can define a transition "vertex formfactors" as off-shell corrections to the "tree" effective couplings. To have these formfactors, like $\rho \pi \pi$, $\omega \rho \pi$ etc., we should collect all needed loop (quark and meson) contributions, which, in general, depend on some NP scale and cannot be considered in the chiral limit (CL) only.  Note, formfactor $\rho q q$ for off-shell quarks has, as it should be, an extra components $\sim k^{\mu}$, where $k$ is the (external) quark momenta. However, as it follows from a preliminary estimations, the value of $\Lambda$ in the diagrams of $\pi q \pi$ and $q \sigma q$ types can be fitted (we get $\Lambda \sim (0.5-0.7) \, \mbox{GeV}$) to provide contributions of the off-shell formfactors into physical parameters (decay widths and so on) no more than $\sim 10 \%$. It means, the gauge principle, i.e. the universality of "tree" effective gauge couplings takes place in the model with this accuracy. 

So, the next order corrections for the diagrams considered can be clearly divided on two types. First, there are a loop corrections which cannot be reduced to contributions into the effective "tree" vertices and off-shell transition formfactors. The role of such loops was studied in \cite{29g}, \cite{29gg}, in particular, for the tree level reactions like $\rho \to \pi \pi \gamma$. 
Second, the rest loop corrections represent on-shell ("tree") and off-shell formfactors with corresponding tensor structure. Approximately, an account of the formfactors can be reproduced by some multiplicative parameter in tree amplitudes. Important, this prescription - an obvious discrimination of an effective tree on-shell constant couplings (remind, the gauge principle demands the same $\rho \pi \pi$ and $\rho q \bar q$ tree couplings) and off-shell formfactors clearly depending on the NP cutoff parameter -  prevents us from the double-counting in the model. Note, in these formfactors masses of internal particles interpret as the physical ones, i.e. all needed self-energy contributions are taken into account.

Therewith, decays like $\omega \to \pi \gamma$, $\rho \to \pi \gamma$, $\omega (\rho) \to 3 \pi$, $\sigma \to \gamma \gamma$ and $\pi^0 \to \gamma \gamma$ are described by finite quark loop diagrams. With a good accuracy, the vertex couplings in these diagrams can be taken as the effective $\rho q \bar q$ ($\rho \pi \pi$) tree constants (not formfactors) in agreement with the experimental data. Note, these arguments can be applied to the electromagnetic vertices too - the vertices are renormalized by strong interactions, but the (formfactor) corrections from quark and meson loops are small in the case under consideration. This fact is endorsed by calculations of loop processes with on-shell photons, in particular, $\pi^0 \to \gamma \gamma$ (see below). Thus, the consistency and supportability of the approach is verified by the results for all types of possible reactions - both at the tree and the loop level.

To fix the set of the gauge physical
couplings and vacuum shifts, we use experimentally observable
mass spectrum and widths of some two-particle hadronic decays of
vector mesons \cite{29d} - in our model these processes occur at the tree level (remind, they define on-shell "tree" couplings). Due to the gauge universality, this approach raises the
predictability of the model when the VMD processes are considered.

 Namely, constants $g_2$, $\sin \phi$ (or $g_1$) and $cos \theta$ are
fixed from the experimental values of $\Gamma
 (\rho^+ \to \pi^+ \pi^0),$ $\Gamma
 (\rho^0 \to \pi^+ \pi^-)$ and  $\Gamma
 (\omega \to \pi^+ \pi^-)$. It should be noted, the last decay
 takes place due to the mixing (\ref{E:11}) only.
Then, the value of $g_{0}$ can be extracted from the relation $e=g_{0}\cdot
 g_2/(g_{0}^2+g_2^2)^{1/2}$.

Then, for the $\rho$ and $\omega$ decays we use tree formulas of the form:
$$\Gamma(V \to \pi_1 \pi_2)=\frac{g_2^2\cdot d_{\theta \phi}}{48\pi}\cdot \frac{1}{m_{V}} \cdot \lambda (m_{\pi_1},m_{\pi_2},m_{V})^{3/2},
$$
Here $V$ is $\rho^0,\,\,\rho^+$ or $\omega$-meson, $d_{\theta \phi}= \cos^2\theta\cos^2\phi$ for the $\rho^0$ decay and $d_{\theta \phi}= \cos^2\theta\sin^2\phi$ for the $\omega$ decay. For the $\rho^{\pm}$ decay $d_{\theta \phi}=1$. $m_V$ is the vector meson mass and
$\lambda (x,y,z)^{1/2}=(x^2+y^2+z^2-2xy-2xz-2yz)^{1/2}$ is the known Kallen function. $\pi_{1,2}$ are $\pi^+,\, \pi^-$ or $\pi^0$ states.
 Thus, from the experimental data and (\ref{E:12c}) we get the consistent set of values of these main parameters,
 which will be used to describe the meson properties in the model:
 \begin{align}\label{E:13}
 g_{0}^2/4\pi = 7.32\cdot 10^{-3}, \,\,\,
 g_1^2/4\pi= 2.86, \,\,\, g_2^2/4\pi = 2.81,\notag\\
 \sin \phi = 0.031, \, \sin \theta = 0.051,\, v_1^2 + v_2^2\approx
 (250.7 \,\mbox{MeV})^2.
 \end{align}

 %In our strategy of calculations strong couplings are extracted from the above mentioned processes as the
% effective final values.
%So, to prevent an analysis from the double counting, we do not
%take into account loop corrections to these couplings. At the same
%time, electromagnetic vertices should be renormalized
% by the strong interactions (for details, see the decay $\rho^0\to e^+e^-$ which is considered in the fourth
% section).
It should be added,
the tree form of the mixing (\ref{E:11}) is caused by the
diagonalization of the real mass matrix (which is generated
by the vacuum shifts without any self-energy
contributions) with the help of the real orthogonal matrix. Having real values of the tree gauge couplings, the
relative phase in the $\rho \pi \pi$ and $\omega \pi \pi$ channels
at the tree level (see (\ref{E:12a})) cannot be reproduced. Experiments, however,
indicate that there is a nonzero relative phase shift between
amplitudes of $e^+e^- \to \rho^0 \to \pi^+ \pi^-$ and $e^+e^- \to
\omega \to \pi^+ \pi^-$ processes (it is interpreted as a
consequence of the $\rho -\omega$ mixing in the pion formfactor).

To introduce the relative $\rho- \omega$ mixing phase in a phenomenological way, physical vector states of $\rho-$ and $\omega-$ mesons are assumed as  a linear combinations of pure isospin states, $|\rho_I>$ and $|\omega_I>$,
 with a complex coefficients
\cite{29e}-\cite{29g}. The phase value is fixed by the experiment data
\cite{29e}-\cite{29k}.
%In our case, the complexity cannot be directly introduced into the superposition of operator fields (\ref{E:11}),
%because it describes the physical neutral fields.
To reproduce the relative phase in the model considered, we can take into account an imaginary parts of meson ($\sigma,\, f_0,\, a_0,\,\pi$) loop corrections to the vertices (even if imaginary parts of the quark loops throw out in the "confinement approximation"). So, we can introduce the needed imaginary part, for example, into the mixing parameters. Because the small relative $\rho-\omega$ phase is inessential for the processes discussed in this paper, their absolute values,$|s_ {\phi}|$ and $|c_{\phi}|$ can be used instead of tree $\sin \phi$ and $\cos \phi$ in the superposition (\ref{E:11}) and also in (\ref{E:12a}) and (\ref{E:12b}). If the phase is wanted for some process, the necessary phase value can be provided due to presence a sufficient number of free parameters in the scalar sector.
%Then, the corresponding phase should be
% introduced into the renormalized couplings as the vertex factors. Namely, in (\ref{E:12a}) and (\ref{E:12b}) parameters $|s_ {\phi}|$ and $|c_{\phi}|$ %occur instead of $\sin \phi$ and $\cos \phi$, correspondingly. This treatment
%is equivalent to the known approach when physical meson states are the complex superpositions of pure isospin vector
%states \cite{29e}-\cite{29g}.
%In particular, the relative phase can arise from
%loop contributions (meson $\sigma,\, f_0,\, a_0,\,\pi-$ loops).
%Certainly, these free parameters should be consistent with scalar meson decays properties.
%An analysis of the vector mesons mass matrix
%with an account of loop contributions and the study of the $\rho -
%\omega$ mixing in the model will be the subject of a forthcoming work.

As to the scalar meson structure, it is
possible to describe the mass spectrum and decay channels of the
scalar mesons  $f_0(600)$ ($\sigma$), $a_0(980)$ and $f_0(980)$ due to presence of a sufficient number of free parameters in the sector.
To illustrate this statement, in addition to (\ref{E:12a}) we also
give a part of the Lagrangian which describes an interaction of
the scalar mesons with $\pi$-mesons:
 \begin{equation}\label{E:14}
  L_{\pi h}=(\pi^0\pi^0+2\pi^{+}\pi^{-})(g_{\sigma\pi}\sigma_0+g_{f\pi}f_0+g_{a\pi}a_0).
 \end{equation}
 This part of the Lagrangian contains free coupling constants and it makes possible to describe the dominant decay
 channels of scalar mesons -
  $f_0(980)\to\pi\pi$ and $\sigma_0\equiv f_0(600)\to \pi\pi$ ~\cite{29d}.
The decay channel $a_0\to\pi\pi$ is not observed and from the model coupling
\[g_{a\pi}=h(v_2-v_1)/\sqrt{2}\] it follows that $v_2\approxeq v_1$.
So, in the scalar sector the residual global $SU(2)$- symmetry approximately takes place after the shift.
Exact symmetry, $v_1=v_2$, is not possible because it forbids the suppressed decay channel $\omega\to\pi^{+}\pi^{-}$,
which is observed experimentally.
  Evidently, the inequality $v_1\ne v_2$, violating the global isotopic $SU(2)$ symmetry,
is related with an account of the electromagnetic interaction in
the model.

Note, an isotriplet structure of scalar fields $a_0$ occurs in the model physical
Lagrangian with an accuracy up to small mixing angles, $\phi$ and
$\theta$. From initial Higgs doublets with $Y=\pm 1/2$ after the
spontaneous breaking, two charged scalar combinations arise. They can
be unified with the one of the residual neutral components to
constitute the isotriplet, which can be identified with the isotriplet $a_0(980)$. To clarify this, we give $a\omega \rho$ and $aa\rho$ parts
of the model Lagrangian for zero mixing angles, $\phi$ and $\theta$
(i. e. $v_1=v_2=v$):

\begin{align}\label{E:14a}
L_{a\omega \rho} &= \frac{1}{\sqrt{2}}\,v g_2 g_3\,\omega^{\mu}(a^-
\rho^{+}_{\mu}+ a^+ \rho^{-}_{\mu}+a^0 \rho^{0}_{\mu})=
\frac{1}{\sqrt{2}}\,v g_2
g_3\,\omega_{\mu}\rho^{\mu}_{\alpha}a_{\alpha},\notag \\
L_{aa\rho} &= \frac{i}{2}\, g_2\, [\rho^{0\mu} (\partial_{\mu}a^+
\cdot a^- -\partial_{\mu}a^- \cdot a^+)+\rho^{+\mu}
(\partial_{\mu}a^-
\cdot a^0 -\partial_{\mu}a^0 \cdot a^-)\notag \\
&+\rho^{-\mu} (\partial_{\mu}a^0 \cdot a^+ -\partial_{\mu}a^+ \cdot
a^0)]= \frac{i}{2}\,
g_2\,\epsilon_{\alpha\beta\gamma}\rho^{\mu}_{\alpha}a_{\beta}\partial_{\mu}a_{\gamma}.
\end{align}
Here we introduced two isovectors, $(\rho^+,\rho^0,\rho^-)$ and
$(a^+,a^0,a^-)$.
As it is seen from the folded invariant forms in (\ref{E:14a}), we get
an exact isotriplet structure of the interaction in the zero mixing
approximation. It is slightly broken by the small mixing with the
singlet vector fields.

Radiative decays of scalars can be analyzed in the model too.
At the tree level two-photon decays of scalar mesons are absent. However, these decays can be considered due to the quark and meson loops. So, the model reproduces the experimental situation in
 these channels (see \cite{29d}). The detailed analysis will be given in a next paper.
 Note, the dominant decay channel $a_0(980)\to \eta \pi$
 can be also described by introduction of an extra term, $\eta \pi_a H^+_A \tau_a H_A$, into
 the initial Lagrangian. This term is admissible by the model symmetry.

As to the interpretation of scalar states in terms of quark
operators, the problem is not solved yet finally. Moreover, it is known
that $f_0(600)$ can be interpreted as a (nearly) gluonium state
(see, for example, \cite{29pa}-\cite{29cd}), and $f_0(980),
\,\, a_0(980)$ are not pure two-quark states. Rather, they are
often interpreted as a four-quark states with an admixture of
gluon and/or $K \bar K-$ components (see \cite{29ka}-\cite{29kg}
and references therein). It is clear that an account of the
strangeness effects (in particular, $a_0 \to K \bar K$ decay) can
not be realized in the frame of the model. To do this, the gauge
symmetry should be extended at least to $SU(3)$ (see also
\cite{18}). However, this problem is beyond the scope of this
paper and will be considered later. To analyze
radiative processes with vector mesons, it is enough to consider
now the physical Lagrangian (\ref{E:12a}) where the "tree" vertices are the effective on-shell constants now. Note that the conception of the gauge and Higgs fields components in the model provides
the invariance of the physical Lagrangian under the $CP$
transformations.

 Because of the model two-level structure (meson-meson and quark-meson),
interactions at these levels are tested independently in
two-particle and three-particle channels. It is seen from
(\ref{E:12a})
 that the meson-meson Lagrangian classifies both tree and loop level processes,
while the quark-meson interactions occur in the model as loop
contributions only.

  \section{Radiative decays $\rho^0\rightarrow\pi^+\pi^-\gamma$ and
  $\omega\rightarrow\pi^+\pi^-\gamma$}

 Radiative decays of neutral vector mesons of the type $V\to \pi \pi \gamma$ are an object of steady attention during some
 decades (see \cite{29a}, \cite{29b}, \cite{29g}, \cite{29gg} and also \cite{30}-\cite{33}). Experimental investigation of these reactions and accompanied theoretical speculations
 contribute to the understanding of hadron
intermediate states and low
 energy dynamic of meson interactions.

In the channels where the
 charged pair $\pi^+ \pi^-$ is formed, dominant contribution comes from the tree diagrams
 corresponding to the vector dominance approximation \cite{29g}, \cite{31}.
Then, radiative decays $\rho^0 \to \pi^+\pi^-\gamma$ and $\omega \to
 \pi^+\pi^-\gamma$ are described by meson-meson sector of the gauge model.

 At the tree level, the former decay is represented by the diagrams in
 Fig.1.
%-------------------------------------------------------------
\begin{figure}[h!]
\centerline{\epsfig{file=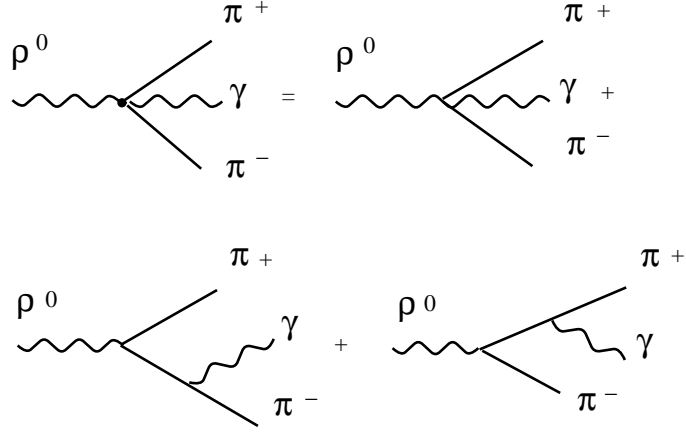,width=9cm}}
\caption{Feynman diagrams for radiative decay $\rho^0\to \pi^+
\pi^- \gamma$.} \label{fig:Feynm1}
\end{figure}
%-------------------------------------------------------------
 The total amplitude for the process is:
 \begin{equation}\label{E:1}
  M^{tot} =\frac{i g e^{\mu}_{\rho}e^{\nu}_{\gamma}}{8
 \pi^2(k^0_{\rho}k^0_{\gamma}k^0_+k^0_-)^{1/2}}\left[g_{\mu \nu} +
 \frac{2k^-_{\mu}k^+_{\nu}}{(k_{\gamma}+k_+)^2-m^2_{\pi}+}+
 \frac{2k^-_{\nu}k^+_{\mu}}{(k_{\gamma}+k_-)^2-m^2_{\pi}}\right].
 \end{equation}
Here $g=e g_2 \cos\theta \cos\phi$ (we do not consider $\rho-\omega$ interference and do not include off-shell formfactors obviously) and $e^{\mu}_{\rho}, \,
e^{\nu}_{\gamma}$ are polarization vectors for $\rho^0$ - meson
and photon, $k_{\rho}, \, k_{\gamma}, \, k_+, \, k_-$ are
4-momenta for all particles in the process. In (\ref{E:1}) we omit
all terms which are equal to zero on the mass shell in the
transversal gauge. For comparison with the experimental spectrum
of photons (see \cite{33}), the differential width is presented in
the form:
\begin{equation}\label{E:2}
d\Gamma(E_{\gamma})/dE_{\gamma} =
\frac{G}{\kappa}\left(F_1(\kappa)+F_2(\kappa) \ln F_3(\kappa)\right),
\end{equation}
where:
\begin{align}\label{E:3}
\kappa &= E_{\gamma}/m_{\rho}, \,\,\, G = \alpha_{em}\cdot g_2^2
\cos^2\theta \cdot \cos\phi^2/24 \pi^2,\,\,\, \mu
=m_{\pi}^2/m^2_{\rho}\,, \notag\\
&F_1(\kappa)=\left(\frac{1-2\kappa-4\mu}{1-2\kappa}\right)^{1/2}
\left(-1+2\kappa+4\kappa^2+4\mu(1-2\kappa)\right);\notag\\
&F_2(\kappa)=1 -2\kappa-2\mu(3-4\kappa-4\mu);\notag\\ &F_3(\kappa)
=\frac{1}{2\mu}\left[1-2\kappa-2\mu+\left((1-2\kappa)\cdot (1-2\kappa -
4\mu)\right)^{1/2}\right].
\end{align}

%-------------------------------------------------------------
\begin{figure}[h!]
\centerline{\epsfig{file=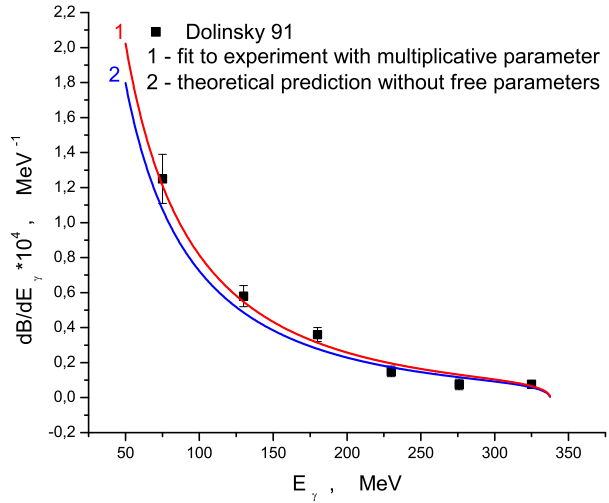,width=9cm}}
\caption{Photons spectrum in $\rho\to 2\pi\gamma$ decay .}
\label{fig:Curve}
\end{figure}
%-------------------------------------------------------------

 Our numerical results, which follow from (\ref{E:2}) and (\ref{E:3}), agree with the results of
\cite{30}-\cite{32} given by the vector dominance approach. In
Fig.2 the theoretical spectrum of photons in comparison with the
experimental data (from ~\cite{33}) is represented. The curve (2)
in Fig.2 describes the spectrum normalized by the total width,
$dB(E_{\gamma})/dE_{\gamma}=1/ \Gamma_{tot}^{\rho} \cdot
d\Gamma(E_{\gamma})/dE_{\gamma}.$ At the tree level only, it does not depend on the
model coupling constants if the total width of $\rho$-meson in the denominator is calculated with the "tree" values of couplings. Here we used, however, the experimental value of $\Gamma_{tot}^{\rho}$, so the normalized branching depends on the $\rho \pi \pi$ and $\gamma \pi \pi$ couplings.

Then, consider the curve (1) representing the model fit of the experimental data with
the help of a single free multiplicative parameter. As it is seen,
this fitted curve (1) improves the description at the photon low
energy range. This improvement can be explained by
an account of $q^2-$ dependence of couplings (i.e. an introduction of corresponding off-shell formfactors with the fixed value of NP parameter $\Lambda$). If we consider these formfactors, and the
spectrum is normalized on the theoretical $\Gamma_{tot}^{\rho}$, the answer depends on the ratio of $g_{\rho \pi \pi}$
formfactors at the different energy scales. More exactly, in the denominator
($\Gamma_{tot}^{\rho}$) this effective coupling is defined on the
mass shell of final pions. At the same time, $g_{\rho \pi \pi}$
for the spectrum $d\Gamma(E_{\gamma})/dE_{\gamma}$ is the "running
coupling" (formfactor) because intermediate pion is off-shell (see Fig.1). A small correction can arise from the electromagnetic $\gamma \pi \pi$- formfactor too. Totally, as it follows from the fit, this multiplicative parameter increases the ratio no more than $(8-10)\%$, due to the formfactors contribution which is approximately $(4-5)\%$.

However, this improvement is ambiguous because of a large
uncertainties in calculated and experimental photon spectrum at
low energies (an accurate account of soft photons is important too).
Note, also, that branching for the decay, which is defined by the
curve (1), agrees with the experiment data worse than the
branching following from the curve (2). To describe the spectrum fine structure
near $E_{max}$, it was suggested to consider the
loop corrections (see \cite{29g}, \cite{30}, \cite{31}), which, certainly, do not contribute to the formfactors.
In any case, this
consideration is reasonable only if the more detailed and reliable
experimental data on the photon spectrum are available.

%It depends on the ratio of $g_{\rho \pi \pi}$ values taken at the
%different energy scales. In the denominator
%($\Gamma_{tot}^{\rho}$) this effective coupling is defined on the
%mass shell of final pions. At the same time, $g_{\rho \pi \pi}$
%for the spectrum $d\Gamma(E_{\gamma})/dE_{\gamma}$ is the running
%coupling because intermediate pion is off-shell (see Fig.1).
Integration of (\ref{E:2}) from $E_{\gamma}^{min} = 50
\,\mbox{MeV} $ up to $E_{\gamma}^{max} = m_{\rho}(1-4\mu)/2$ gives
the value of partial $\rho$- meson branching $B(\rho^0\to \pi^+
\pi^- \gamma) = 1.17 \cdot 10^{-2}$ which slightly exceeds the
experimental value $B^{exp}(\rho^0\to \pi^+ \pi^- \gamma) = (0.99
\pm 0.16)\cdot 10^{-2}$ \cite{29d}. An account of the loop
contributions (with the phenomenological couplings) leads to the
result: $B^{phen}(\rho^0\to \pi^+ \pi^- \gamma) = (1.22 \pm
0.02)\cdot 10^{-2}$ ~\cite{31}. It should be noted that there is
some discrepancy between the experimental branching and the one
following from the integration of spectrum (see Fig.2). Namely, an
excess of $B^{theor}$ over $B^{exp}$ can be caused by the
deviation of the theoretical spectrum from the experimental one at
the energy $E_{\gamma}\sim 100\, \mbox{MeV}$, where there are
large uncertainty, and at energy range $E_{\gamma}<75
\,\mbox{MeV}$, where the measurements are absent. So, to test a
more detailed model predictions we need in a more precise
experimental data.

Decay characteristics of the process $\omega\to \pi^+ \pi^-\gamma$
at the tree level are computed analogously with the following
replacement in (\ref{E:1}) - (\ref{E:3}): $ \cos\phi \to \sin\phi$
in $G$, and $m_{\rho} \to m_{\omega}$ in $\kappa$ and $\mu$. The
partial width for the decay is damped by the small mixing
parameter, $\sin\phi \approx 0.034$, so we have $B(\omega\to
\pi^+ \pi^- \gamma) = 4.0 \cdot 10^{-4}$ and $B(\omega\to \pi^+
\pi^- \gamma) = 2.6 \cdot 10^{-4}$ for $E^{min}_{\gamma} = 30 \,
\mbox{MeV}$ and $50\, \mbox{MeV}$, respectively. These estimations
do not contradict to the experimental restriction
$B^{exp}(\omega\to \pi^+ \pi^- \gamma) \le 3.6 \cdot 10^{-3}$
~\cite{29d} and agree with the theoretical results of ~\cite{29g}.

Thus, the same set of the fixed parameters (the gauge "tree" constant and
two mixing angles) describes two different decays which arise at the (tree)
level. Certainly, loop corrections can be important for the
$\omega\to \pi^+ \pi^- \gamma$ decays due to the smallness of the
tree contribution (see, for example ~\cite{29g}, \cite{29gg}). In an analogy
with the results of these papers, loop corrections can increase
$B(\omega\to \pi^+ \pi^- \gamma)$ up to $(2-3) \cdot 10^{-3}$,
which does not contradict again to the upper limit of the
experiment.

As it was noted above, an account of the loop corrections to the
decay $\rho^0\to \pi^+ \pi^- \gamma$ tree  width (see
\cite{29g} and \cite{31}-\cite{33}) was intended to describe the photon spectrum
fine structure. However, it increases the discrepancy between the
model and experimental values of the total width. Quantitatively,
this effect depends on the underlying model. Moreover, in
effective theories an account of loop diagrams (for the tree level processes) has some additional subtleties connected with
the compensation of divergencies and renormalizability. For the
processes which take place at the loop level only, these problems are
absent --- all divergencies are summed to zero when all external
lines are on the mass shell \cite{36}. An examples of such loop
processes will be given in the next section.

The usage of parameters from (\ref{E:13}) together with the
constants $g_{\rho ee}$ and $g_{\omega ee}$ extracted from the
widths of $\rho^0 \to ee$ and  $\omega \to ee$ decays, leads to
the correct estimation of the cross section in the peak vicinity,
$\sigma^{theor} \approx \sigma^{exp}\approx 1.3 \,\mbox{mkb}$. As
it was noted above, the tree approximation
%(without an account of
%the $\rho -\omega$ mixing at the loop level)
does not describe the resonance curve in this region in details. For the detailed consideration, a relative
$\rho-\omega$ phase should be included, but it is beyond the scope of this paper.

\section{Loop radiative
%and leptonic
decays of mesons}
%\subsection{$\rho^0\rightarrow\pi^0\gamma$, $\omega\rightarrow\pi^0\gamma$ decays}
In the most papers these decays are defined by the
phenomenological vertices which are introduced at the tree level
\cite{1}. In the quark-meson model, radiative decays $\rho^0,\,
\omega \to \pi^0\gamma$ and three-particle decays $\omega, \rho
\to 3 \pi$ occur via quark loops only with the gauge (tree) vertices (see also \cite{27b}, \cite{28}, \cite{38a}). Remind, corresponding formfactors depending on some NP parameter can be add to the tree couplings. Then, if the formfactors had introduced, loop corrections of another type should be considered too.

Corresponding one-loop diagrams for the decays $\omega,\, \rho^0 \to
\pi^0\gamma$ are shown in Fig.3, where $q = u,d$ denote quark
fields.
%-------------------------------------------------------------
\begin{figure}[h!]
\centerline{\epsfig{file=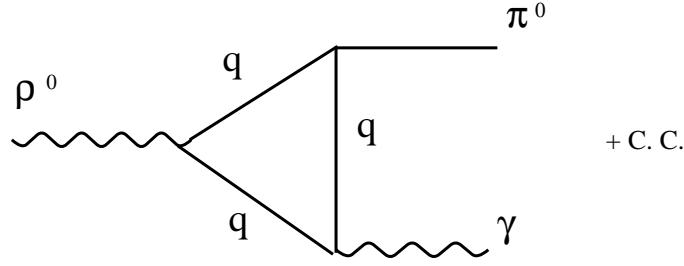,width=9cm}}
\caption{Feynman diagrams for the radiative decay $\rho \to\pi^0
\gamma$.} \label{fig:Feynm2}
\end{figure}
%-------------------------------------------------------------

Total amplitude for  the
process $\omega \to \pi^0\gamma$ has the following form
\begin{equation}\label{E:4}
M_{\omega}=\frac{-2i\pi^2 N_cgm_q}{(2\pi)^{9/2}(2p^0k^0_{\gamma}k^0_{\pi})^{1/2}}
e^{\mu}_{\omega}e^{\nu}_{\gamma}k^{\alpha}_{\gamma}p^{\beta}\epsilon_{\mu
\nu \alpha \beta} \cdot C_0(0,m_{\omega}^2,m_{\pi}^2;m_q,m_q,m_q).
\end{equation}
Here $N_c = 3$ is the color factor, vertex constant $g = g_1e
\varkappa \cos\phi$ ($\varkappa$ is the constant of
$\sigma q\bar q$ and $\pi q \bar
q$ interaction, and the Goldberger - Treiman relation defines the coupling as $\varkappa=m_q/f_{\pi}$). For the constituent quark mass we
suppose $m_u \approx m_d = m_q$. $C_0(0,
m_{\pi}^2,m_{\omega}^2;m_q,m_q,m_q)$ is here the special case of the three-point Passarino
- Veltman function \cite{37}.
In a general case for all incoming momenta:
$$
C_0(p_1^2,p_2^2,Q^2;m_1,m_2,m_3)=\frac{i\mu^{4-n}}{\pi^2}\int \frac{d^n q}{d_0 d_1 d_2},
$$
where $d_0=q^2-m_1^2+i\epsilon$, $d_1=(q+p_1)^2-m_2^2+i\epsilon$, $d_2=(q+Q)^2-m_3^2+i\epsilon$ and $Q=p_1+p_2$.

Using an obvious form of the $C_0$
in (\ref{E:4}), we have:
\begin{equation}\label{E:5}
\Gamma(\omega \to \pi^0 \gamma) =\frac{3\alpha
g_1^2}{2^7\pi^4}\cos^2\phi\,\,
m_q\frac{m_q^3}{m_{\omega}f_{\pi}^2}\left(1-\frac{m_{\pi}^2}{m_{\omega}^2}\right)
|L_{\omega}|^2.
\end{equation}
Function $L_{\omega}$ occurs from the $C_0$ function
$$L_{\omega}=Li_2\left(\frac{2}{1+\sqrt{\lambda_1}}\right)+Li_2\left(\frac{2}
{1-\sqrt{\lambda_1}}\right)-Li_2\left(\frac{2}{1+\sqrt{\lambda_2}}\right)
-Li_2\left(\frac{2}{1-\sqrt{\lambda_2}}\right),$$ where $\lambda_1
= 1 -4m_q^2/m_{\omega}^2, \, \lambda_2 = 1 -4m_q^2/m_{\pi}^2$ and
the decay constant $f_{\pi} = 93\,\mbox{MeV}$. In (\ref{E:5}) it
was used again the Goldberger-Treiman relation $\varkappa \approx
m_q/f_{\pi}$. Then, the
constituent quark mass value, which is taken here as a free parameter,
can be found from the widths fit.

The decay $\rho^0 \to \pi^0\gamma$  is described by the diagrams in
Fig.3 with the corresponding replacement of the coupling constant. The
expression for the width is
\begin{equation}\label{E:6}
\Gamma(\rho^0 \to \pi^0\gamma) =\frac{\alpha g_1^2}{3\cdot
2^7\pi^4}\cos^2\phi\cdot\left(\cos\theta\cdot\frac{g_2}{g_1}\right)^2
m_q\frac{m_q^3}{m_{\rho}f_{\pi}^2}\left(1-\frac{m_{\pi}^2}{m_{\rho}^2}\right)
|L_{\rho}|^2.
\end{equation}
 Note that the relation $\Gamma(\rho^0 \to \pi^0\gamma)/\Gamma(\omega \to \pi^0\gamma) \approx 1/3^2$ follows
from the isotopic structure of $\rho qq$, $\omega qq$ and $\gamma
qq$ vertices (\ref{E:12a})-(\ref{E:12b}) (here we omit small terms
proportional to $\sin\theta$ and $\sin\phi$). Both the widths are
in a good agreement with the experimental data \cite{29d} when the
effective quark mass value is $m_q = 175 \pm 5 \,\mbox{MeV}$:
\begin{align}\label{E:7}
 \Gamma^{theor}(\omega \to \pi^0\gamma)=0.74\pm
0.02 \, \mbox{MeV},\quad \quad \Gamma^{exp}(\omega
\to \pi^0\gamma)=0.76 \pm 0.02 \,\mbox{MeV};\notag \\
\Gamma^{theor}(\rho^0 \to \pi^0\gamma) = 0.081\pm 0.003 \, \mbox{MeV},
\quad \quad \Gamma^{exp}(\rho^0 \to \pi^0\gamma)=0.090 \pm 0.012 \,
\mbox{MeV}.
\end{align}
Thus, the fitting of two widths by one model parameter - mass of the constituent quark $m_q$,
gives $m_q=175 \pm 5\,\mbox{MeV}$. Remind, we find the widths not in the limit of low (zero) external lines (see, for instance, \cite{28}), but calculate all integrals exactly.

Note, this value of $m_q$ is fixed from the fit if we keep nonzero imaginary part of these amplitudes. It means, we suppose the constituent quarks can create and propagate in a NP vacuum domain. In essence, instead of the known approximation of the "naive" confinement (when the imaginary part of quark loop equals zero "by hand" - see \cite{37a} and, for example, \cite{27b}, \cite{28}), we suggest another "naive" idea: inside of the NP medium real constituent quarks are created by a vector or a (pseudo) scalar current and as a free states they run to a distances $\sim 1/\Lambda$ before annihilation. So, these quarks cannot be an asymptotic states in the model. The $\Lambda$ defines a scale where the NP medium is capable to transform the (real) quark pair into the meson. Possibly, the scale $\Lambda \sim m_{\sigma}$ because the $\sigma$-meson simulates a scalar "gluonium", which effectively links quarks together in the model. The same scale $\Lambda$ should define the diverged parts of the vertices formfactors mediated by the quark and meson loops. As it is seen, the obtained quark mass is significantly lower than the standard value $m_q \approx 300 \, \mbox{MeV}$. The reason is the "real" and relatively light constituent quarks attractively interact with each other, exchanging by "gluoniums" (virtual scalar and pseudoscalar meson states). The last ones not only contribute into the quark mass via self-energy diagrams and the quark binding at the large distances ($\sim 1/\Lambda$), but also exist outside the hadron vacuum as the real mesons. In other words,
a constituent quark contains only a part of a hadron gluon component,
and true gluon core inside of a NP vacuum is simulated by the "gluonium".
%Outside of the hadron, it manifests itself as a meson state.
So, the imaginary parts of amplitudes describe the constituent quark propagation up to the distances, where the NP effects are sufficiently large to bind the quarks into a real meson. Attractive forces between the quarks lead to the effective decreasing of their mass.

%\subsection{Comment to the constituent quark mass value in the model}
%The obtained  quark mass value can be understood also in a connection
%with the schematic representation of the quark-gluon content of
%hadrons.
Schematically, for the nucleon and $\rho$- meson masses we
can write approximately
\begin{equation}\label{E:15}
 m_N=3m_q+m_G;\,\,\,m_{\rho}=2m_q+m_G.
\end{equation}

Here,$m_q$ is an effective quark mass and $m_G$ is the mass of the same
internal "gluon" component of meson and baryon structure, as it is supposed here.
Instead of explicit fundamental gluon fields, in the effective model
there is $\sigma$- meson, which can be interpreted as an effective
scalar "gluionium" state with a mass close to $m_G$.

From this system of equations it follows $m_G\approx
435\,\mbox{MeV}$ and $m_q\approx 170\,\mbox{MeV}$. The latest
value nearly coincides with the quark mass assessment resulting
from the decay channels analysis. Note that quark mass value
$m_q\approx 230 \,\mbox{MeV}$ and the $m_{\sigma}\approx
470 \, \mbox{MeV}$ result from the composite-meson model
with the four-quark interaction \cite{38aa}.

%So, the model discussed introduces the constituent quark which is
%differed from the traditional effective quark component of a
%hadron with $m_q=m_N /3\approx 300\,\mbox{MeV}$ (see, however,
%\cite{38a}).

The value of $m_G$ makes it possible to interpret the
meson $\sigma_0=f_0(600)$ as the (mostly) glueball,
nonperturbative state \cite{38b,38c}. This interpretation is
consistent with an ideas of the gluon nature of $\sigma-$ meson
(see ~\cite {38} and references therein for the review of scalar
meson properties).
%If it is the case, the low value of the
%constituent quark mass can be treated as the consequence of a
%model separation of the quark and "gluon" degrees of freedom in a
%hadron. Hence, the value of $m_q$, which is essentially lower than
%$300\,\mbox{MeV}$, provides the agreement of the model results
%with the experiment for the processes under consideration (for the case the imaginary part of corresponding amplitudes should be taken into account). Of
%course, if we explanate the $\sigma$-meson as the effective
%glueball state, we need in an detailed analysis of its properties
%and manifestations in the model. This studying will be presented
%in a forthcoming work.
Note, the $\sigma-$ meson was also
carefully analyzed supposing it is a $\bar qq$ state, in
~\cite{39,40}, for instance.

Nevertheless, if we use the "naive" confinement approximation of \cite{37a}, \cite{38a} and set the imaginary parts of quark loops zero, from our loop calculations with the tree vertices (not formfactors) we get $m_q \approx (280-300) \, \mbox{MeV}$ in agreement with the known results of quark-meson models. An account of the formfactors instead of tree values can increase the quark mass no more than $\sim 10 \%$, as we estimate.

Certainly, the possibility to keep nonzero imaginary part should be verified for other quark loop amplitudes, for example, for the $\rho \omega \pi$-formfactor. In particular, from the quark triangles analogous to that describing $\pi \gamma \gamma$ process, in the low external momenta approximation (see \cite{27a}, \cite{28}) we get the same result as in \cite{28}. However, if we integrate the vertex exactly, the results in a various kinematical regions for the external momenta differ from this estimation substantially (details will be given in a next paper).

Now let's consider the known decay $\pi^0 \to \gamma \gamma$, which
is described by finite triangle quark diagrams too. Direct calculation leads to the following expression for
the partial decay width:
\begin{equation} \label{E:15a}
\Gamma(\pi^0\to \gamma \gamma)=\frac{\alpha^2 \varkappa^2}{16
\pi^3}m_{\pi}^3m_q^2 [C_0(0,0,m_{\pi}^2;m_q,m_q,m_q)]^2,
\end{equation}
where again $\varkappa=m_q/f_{\pi}$. $C_0(0,0,m_{\pi}^2;m_q,m_q,m_q)-$ is another special case of the scalar three-point
Passarino-Veltman function.
Here the function is reduced to the form
$$
C_0(0,0,m_{\pi}^2;m_q,m_q,m_q)=
\frac{1}{m_{\pi}^2}[Li_2(\frac{2}{1+\sqrt{\lambda}})+Li_2(\frac{2}{1-\sqrt{\lambda}})],
$$
where $\lambda=1-4m_q^2/m_{\pi}^2$. Then, from
Eq.(\ref{E:15a}) it follows numerically
$$
\Gamma(\pi^0\to \gamma \gamma)=8.48 \, \mbox{eV}
$$
for $m_q=175\,\mbox{MeV}$. If the quark mass value $m_q=300 \,
\mbox{MeV}$, the width is $\Gamma(\pi^0\to \gamma \gamma)=7.91 \,
\mbox{eV}$. In an analogous consideration \cite{22}, \cite{41}, the decay
width in the exact chiral limit (corresponding to axial anomaly term)
has the value: $\Gamma(\pi^0\to \gamma \gamma)=7.63 \, \mbox{eV}$.
Note that from Eq.(\ref{E:15a}) we get the same value in the
chiral limit for the $C_0$- function. The experimental width lies
in the interval $7.22 \, \mbox{eV}\le \Gamma(\pi^0\to \gamma
\gamma)\le 8.33 \, \mbox{eV}$ \cite{29d}.

There are some essential aspects of this result. First, in the chiral limit, supposing a $100 \%$ accuracy for the Goldberger-Treiman relation, the width calculated does not depend on the quark mass, and our result is coincide with the width from the paper \cite{22}. Second, exact integration of the amplitude allows, as it would seem, to prefer the value $m_q\sim 300\,\mbox{MeV}$, which provides a better
agreement with the experimental data. Some
important factors, however, should be taken into account for more
accurate analysis: a) a possible violation of the Goldberger-Treiman relation can be as much as $(3-4) \%$ \cite{42}-\cite{42c} (it means
some weak $q^2-$ dependence of $\varkappa$); b) as it was shown
in \cite{43}, the $\pi^0 - \eta-$ mixing can be noticeable for
the case and c) loop corrections to
$\gamma q \bar q$ vertex, i.e. electromagnetic formfactor for off-shell quarks can also be significant in this case.

For
example, taking the quark mass equals $175 \, \mbox{MeV}$ and decreasing effective $\pi q \bar q$ coupling by $4 \%$ only, for the
decay width we get $\Gamma(\pi^0\to \gamma \gamma)=7.82 \,
\mbox{eV}$ also in a good agreement with the experiment. On the other hand, to make the theoretical width consistent with the data, it is sufficient to
decrease the effective electromagnetic coupling  $\gamma q \bar q$ no more than $2\%$. Certainly, the smallness of these contributions into the electromagnetic quark formfactor is provided by the specific value of the NP scale, $\Lambda$ for this formfactor, which is defined by the diverged triangles  of $q \sigma q$ and $\pi q \pi$ types. This value of $\Lambda\approx (0.6-0.8)\,\mbox{GeV}$ correlates with the NP scale, which fixes formfactors of other effective interactions ($\rho \pi \pi$, $\rho q \bar q$).

And third, in this decay an imaginary part of the amplitude equals zero exactly. So, for the case we do not need in the "naive" quark confinement approximation \cite{37a}, \cite{22}: to drop imaginary part of the quark loop amplitude.
%Analogously, in the loop three-particle decays $\omega (\rho) \to
%\pi^0 \pi^+ \pi^-$ analysis
% the ratio $m_{\pi}^2/ m_q^2$ is non-negligible (cf. with
% ~\cite{28} where all loop integrals were computed in the chiral limit).
% Nevertheless, an accurate calculations give for $\Gamma(\omega (\rho)\to 3\pi)$ the
% values which are in agreement with the
% data (these results will be
%discussed in the forthcoming work too).

\section{Conclusions}

The linear sigma-model is the most popular approach for the
description of low-energy hadron interaction. We have considered
the gauge generalization of this model and include (effective)
quark degrees of freedom explicitly. In the gauge scheme suggested
the vector meson dominance takes place in the tree-level
processes. The quark-meson sector describes some loop-level
processes, when quark structure of mesons plays a noticeable role.

The analysis of the model discussed is not completed yet, in particular, in the scalar
sector, which is formed by the Higgs degrees of freedom. We demonstrate a principal possibility to introduce scalar
singlets $f_0(600), \,\, f_0(980)$ and triplet $a_0(980)$ into the
gauge scheme. The structure of the scalar sector and the study of the scalar mesons decays is in preparation now. Certainly, more detailed description of scalar
mesons properties should be realized in $SU(3)$ extension. To account for the chiral partners of vector mesons, we
need in the localization of the total chiral group. However, the processes, which are considered here, can be analyzed without these extensions.

Using an above discussed strategy of the model parameters treatment, we estimate the vertex couplings, avoiding the double counting and keeping the gauge symmetry. Then, the model was applied to some radiative decays
of mesons which are intensively discussed in the literature.
The decays $\rho^0\rightarrow\pi^+\pi^-\gamma$ and
$\omega\rightarrow\pi^+\pi^-\gamma$ have been considered within
the framework of the VMD gauge scheme. The results are in good
accordance with the experimental data. The radiative decays
$\rho^0\rightarrow\pi^0\gamma$, $\omega\rightarrow \pi^0\gamma$ and $\pi^0\rightarrow \gamma \gamma$
are defined by the quark-meson interaction entirely, i.e. they
take place at the loop level only. For these cases we get results
in agreement with the experiment too. The quark mass value and "confinement" at the level of effective quarks are discussed.

From the analysis we fulfilled, it follows that the radiative
decay rates of vector mesons can be calculated with a good
accuracy within the framework of the gauge quark-meson model. In
other words, quantum field approach to the VMD, supplemented by
the quark-meson interaction, is a gauge generalization of the
linear sigma-model.

\end{document}